\documentclass{PoS}

\newcommand{\be}{\begin{equation}}
\newcommand{\ee}{\end{equation}}
\newcommand{\bea}{\begin{eqnarray}}
\newcommand{\eea}{\end{eqnarray}}
\newcommand{\bi}{\begin{itemize}}
\newcommand{\ei}{\end{itemize}}
\newcommand{\ben}{\begin{enumerate}}
\newcommand{\een}{\end{enumerate}}
\newcommand{\bt}{\begin{tabbing}}
\newcommand{\et}{\end{tabbing}}
\newcommand{\Vus}{|V_{us}|}

\title{
   \begin{picture}(0,0)(0,0)%
   \put(355,75){\makebox(0,0)[l]{\textnormal{\normalsize KEK-CP-201}}}%
   \end{picture}%
   Theoretical progress on $|V_{us}|$ on lattice
}

\ShortTitle{Theoretical progress on $\Vus$ on lattice}

\author{Takashi Kaneko
        \\
        \\
        High Energy Accelerator Research Organization (KEK),
        Ibaraki 305-0801, Japan 
        \\
        School of High Energy Accelerator Science,
        The Graduate University for Advanced Studies (Sokendai),
        Ibaraki 305-0801, Japan
        \\
        E-mail: \email{takashi.kaneko@kek.jp}}

\abstract{Recent lattice studies on (semi-)leptonic
          kaon decays towards a precise determination of $\Vus$
          are reviewed. 
          Attention is given to recent unquenched calculations
          and consistency of their results with 
          chiral perturbation theory.
         }

\FullConference{KAON International Conference\\
                May 21-25 2007\\
		Laboratori Nazionali di Frascati dell'INFN, Rome, Italy}

\begin{document}

\section{Introduction} 


Accurate knowledge of the Cabibbo-Kobayashi-Maskawa (CKM) matrix element 
$\Vus$ is important because it gives the basic parameter $\lambda$ 
in the Wolfenstein parametrization of the CKM matrix
and is relevant to a stringent test of CKM unitarity
$|V_{ud}|^2+\Vus+|V_{ub}|^2\!=\!1$.
%
The $K_{l2}$ and $K_{l3}$ decays provide two precise determinations of $\Vus$,
where their dominant uncertainty originates from 
theoretical evaluations of hadronic matrix elements, 
namely $f_K/f_\pi$ and $K_{l3}$ form factors.


Lattice QCD can provide a non-perturbative estimate of these matrix elements
from first principles.
Due to the limitation of the computational resources, however,
some simulation parameters have to be different from those of the real world.
The use of finite lattice spacing $a$ and spatial extent $L$ is 
unavoidable but its effects can be systematically reduced.
It is assumed in the simulations reviewed in this article
that up and down sea quarks are degenerate.
The use of relatively heavy masses $m_{ud, \rm sim}$ 
for degenerate up and down quarks is much more problematic, 
because it could cause a large uncontrolled error 
by extrapolating lattice results to the physical mass $m_{ud}$.
It is therefore vital to simulate quark masses $m_{ud, \rm sim}$, 
where chiral perturbation theory (ChPT) can be safely used as a 
guide for the chiral extrapolation.

The staggered fermions are known to be computationally inexpensive 
\cite{CPUtime:KS},
and led to a precise determination of $f_K/f_\pi$ 
\cite{fK_fpi:Nf3:ImpG+ImpKS:MILC:2004}
from the MILC collaboration's simulations 
at $m_{ud, \rm sim} \! \gtrsim \! m_s/10$ 
\cite{Spectrum:Nf3:ImpG+ImpKS:MILC}.
Their complicated flavor structure is however
a serious obstacle to extensive calculations of 
more involved matrix elements, 
such as the $K_{l3}$ form factors.
While simulations with other discretizations were limited to 
relatively heavy quark masses, typically $m_{ud, \rm sim} \! \gtrsim \! m_s/2$,
at the time of the last conference KAON 2005,
recent algorithmic improvements now enable us to explore 
$m_{ud, \rm sim}$ comparable to that in the MILC's study.
In any lattice studies, consistency between their data and ChPT
is a crucial issue for a reliable chiral extrapolation.



In this article,
we first review recent progress on $f_K/f_\pi$ in Sec.~\ref{sec:fK_fpi},
focusing on the latest update on the MILC's estimate 
and the status of studies with other discretizations.
Section~\ref{sec:kl3} is devoted to the $K_{l3}$ form factors. 
We outline the calculation method and discuss 
the associated systematic errors.
%
%
Finally, our conclusions are given in Sec.~\ref{sec:summary}.


\section{$f_K/f_\pi$} 
\label{sec:fK_fpi}


As pointed out in Ref.\cite{Vus:Kl2+pil2},
$\Vus$ can be extracted from $K_{l2}$ and $\pi_{l2}$ decays 
through the ratio of their decay rates
\bea
   \frac{\Gamma(K \to l \bar{\nu}_l)}{\Gamma(\pi \to l \bar{\nu}_l)}
   & = &
   \frac{\Vus^2}{|V_{ud}|^2}
   \frac{f_K^2}{f_\pi^2}
   \frac{M_K\,(1-m_l^2/M_K^2)^2}{M_\pi\,(1-m_l^2/M_\pi^2)^2}\,
   \left\{1+\frac{\alpha}{\pi}(C_K-C_\pi)\right\}.
   \label{eqn:fK_fpi:Kl2_pil2}
\eea
Radiative corrections parametrized by $C_{K,\pi}$ and 
the muonic decay rates lead to an uncertainty of $\lesssim \! 0.2$\%
in $\Vus$.
The determination of $|V_{ud}|$ from super-allowed nuclear $\beta$ decays
is accurate at the impressive level of 0.05\%.
Therefore the main uncertainty in $\Vus$ 
comes from the theoretical input $f_K/f_\pi$.

Lattice QCD can, in principle, give a precise estimate of $f_K/f_\pi$, 
since each decay constant is calculated from simple (and hence less noisy) 
two-point functions and 
uncertainties due to the lattice scale and renormalization are canceled 
in the ratio.
The dominant error arises from the continuum and chiral extrapolations.
%
%
The original estimate $\Vus\!=\!0.2236(30)$ in Ref.\cite{Vus:Kl2+pil2} 
was obtained with the MILC's estimate 
$f_K/f_\pi\!=\!1.201(8)(15)$ \cite{fK_fpi:Nf3:ImpG+ImpKS:MILC:2004}
from their simulations using an improved staggered action
(the so-called AsqTad action)
at two lattice spacings $a\!=\!0.09$ and $0.12$~fm
and with quark masses down to $m_{ud, \rm sim} \sim m_s/10$ 
\cite{Spectrum:Nf3:ImpG+ImpKS:MILC}.

\subsection{Update on result from staggered fermions}


The MILC collaboration has been steadily updating their simulations.
One of the main improvements in their latest report is that 
their simulations are extended to finer ($a\!=\!0.06$~fm) 
and coarser lattice spacings ($a\!=\!0.15$~fm) 
\cite{fK_fpi:Nf3:ImpG+ImpKS:MILC:2006:1,fK_fpi:Nf3:ImpG+ImpKS:MILC:2006:2}.


It should be noted that simulations with the staggered quark action have
the following theoretical and technical complications.
By construction, a single staggered field describes four species of quark.
This degree of freedom is called ``taste''.
In simulations with single flavor (two degenerate flavors) of quarks,
gauge configurations are generated by taking the fourth (square) 
root of the quark determinant in the Boltzmann weight,
{e.g.}
\bea
   Z_{N_f=1} 
   & = &
   \int [dU] \det [ D ]^{1/4} \exp[-S_g],
   \label{eqn:fK_fpi:rooting}
\eea
where $D$ is the Dirac operator for the {\it four-taste} staggered quark,
$S_g$ is the lattice gauge action of choice, 
and $dU$ represents the path integral over the gauge fields.
It is still actively debated whether
the non-local Dirac operator corresponding to the rooted determinant 
leads to the correct continuum limit
\cite{KS:rooting}.
In addition, the explicit taste symmetry breaking at finite lattice spacings
makes calculations of matrix elements complicated.


In the MILC's latest analysis
\cite{fK_fpi:Nf3:ImpG+ImpKS:MILC:2006:1,fK_fpi:Nf3:ImpG+ImpKS:MILC:2006:2},
they fit the quark mass and lattice spacing dependence of 
the pseudo-scalar meson masses and decay constants
simultaneously using formulas from the so-called staggered ChPT \cite{SChPT},
where effects due to the taste symmetry breaking 
are taken into account.
Analytic and chiral logarithmic terms at NLO 
and a part of analytic terms up to NNNLO 
are included into their fitting function
to achieve a reasonable value of $\chi^2/{\rm dof}$.
Their two observations
increase the reliability of their chiral and continuum extrapolations:
\bi
      \item their fit curve in the continuum limit
            exhibits a curvature towards the chiral limit
            as expected from NLO ChPT;
      \vspace{-2mm}
      \item they obtain low-energy constants (LECs)
            $L_4\!=\!0.1(4)$ and $L_5\!=\!2.0(4)$,
            which are consistent with a phenomenological estimate
            $L_4\!=\!0.0(8)$ and $L_5\!=\!2.3(1)$
            \cite{L4_L5}.
\ei
\vspace{2mm}



They obtain their latest estimate
\bea
   \frac{f_K}{f_\pi}
   & = &
   1.208(2)(+7/\!-\!14),
   \label{eqn:fK_fpi:MILC}
\eea
where the first error is statistical and the second is systematic,
and obtain $f_\pi\!=\!128.6(0.4)(3.0)$~MeV and $f_K\!=\!155.3(0.4)(3.1)$~MeV, 
which are in good agreement with experiment.
The statistical error is remarkably reduced from 
their previous estimate in Ref.\cite{fK_fpi:Nf3:ImpG+ImpKS:MILC:2004}.
The uncertainty in $f_K/f_\pi$ is now dominated by the systematics 
of the combined chiral and continuum extrapolation,
which might be difficult to improve drastically
without extending their simulations to a much wider range of $m_{ud, \rm sim}$ 
and $a$.
Independent calculations with different fermion discretizations 
are highly required 
to reduce the systematic uncertainties 
and to confirm that 
the rooted staggered theory has the correct continuum limit.


\subsection{Status of studies with other discretizations}


Unquenched simulations with other fermion discretizations also 
have a long history, 
leading up to recent studies with the following fermion actions:


\vspace{-2mm}
\bi
   \item (improved) Wilson fermions 

         This traditional formulation is computationally cheap 
         compared to chiral fermions (see below) 
         and useful to simulate large and fine lattices. 
         Its main drawback is the explicit chiral symmetry breaking
         at finite lattice spacing,
         which may distort the chiral behavior of the decay constants 
         \cite{WChPT}.
         The clover action is an improved formulation by removing 
         leading $O(a)$ discretization errors. 
         These discretizations are employed in recent simulations 
         in Refs.\cite{Nf2:Plq+Wlsn:CERN,Nf2:Plq+Clv:QCDSF+UKQCD,Nf3:RG+Clv:PACS-CS}.

   \vspace{-2mm}
   \item twisted mass Wilson fermions 

         This is a variant of the Wilson fermions with the so-called
         twisted mass term \cite{tmQCD},
         which simplifies the mixing pattern in the renormalization 
         of lattice operators remarkably 
         with computational costs
         comparable to Wilson fermion simulations.
         This mass term, however, leads to the explicit breaking 
         of parity and isospin symmetry. Its effects have to be studied 
         carefully, as in large-scale simulations 
         by the ETM collaboration \cite{Nf2:Sym+tmW:ETM}.

   \vspace{-2mm}
   \item chiral fermions

         With the five dimensional domain-wall formulation \cite{DWF},
         chiral symmetry is restored in the limit of infinitely large 
         size $L_s$ in the fifth dimension. 
         It is however $L_s/a$ times costly 
         with respect to the above mentioned Wilson-type fermions.    
         The RBC/UKQCD collaborations simulate three-flavor QCD
         with $L_s/a \! = \! 16$, which leads to the additive quark 
         mass renormalization of a few MeV
         \cite{Nf3:RG+DW:RBC+UKQCD}. 
         The (four dimensional) overlap fermions \cite{overlap} are 
         even more computationally demanding.
         However, it has almost exact chiral symmetry 
         and hence is useful for phenomenological applications
         where chiral symmetry plays an important role.
         The JLQCD collaboration has started large scale 
         simulations in two-flavor QCD \cite{Nf2:RG+Ovr:JLQCD}.
\ei

\FIGURE{
   \centering
   \includegraphics[angle=0,width=0.45\linewidth,clip]{MPSsea_unquenched.eps}
   \vspace{-3mm}
   \caption{
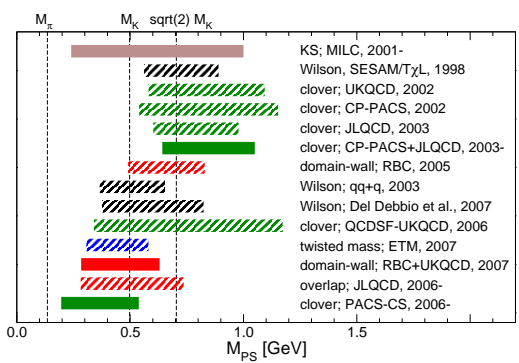
      Region of pion mass simulated in large-scale unquenched calculations
      in two-flavor (shaded band) and three-flavor QCD (solid band).
   }
   \label{fig:fK_fpi:Mpi2}
}


The simulation cost for the above formulations with the commonly used 
Hybrid Monte Carlo (HMC) algorithm \cite{HMC} rapidly increases as
 $m_{ud, \rm sim}$ decreases:
it scales as $\propto m_{ud, \rm sim}^{-3}$ \cite{Berlin_Wall}.
This is why 
previous simulations on relatively large and fine lattices 
\cite{Nf2:Plq+Wlsn:SESAM+TchiL,Nf2:Plq+Clv:UKQCD,Nf2:RG+Clv:CP-PACS,Nf2:Plq+Wlsn:qq+q,Nf2:Plq+Clv:JLQCD,Nf2:DBW2+DW:RBC,Nf3:RG+Clv:CP-PACS+JLQCD}
are limited to heavy quark masses 
typically  $m_{ud, \rm sim} \! \gtrsim m_s/2$,
as shown in Fig.~\ref{fig:fK_fpi:Mpi2}.
However, recent algorithmic improvements \cite{algorithm1,algorithm2}
enable us to simulate much smaller values of $m_{ud, \rm sim}$,
which are now comparable to those in the MILC's simulation with 
the staggered fermions.


In Fig.~\ref{fig:fK_fpi:fPS_vs_Mpi2}, 
we plot the pion decay constant obtained with Wilson-type and chiral fermions
\cite{Nf2:Sym+tmW:ETM,Nf2:RG+Ovr:JLQCD,Nf2:Plq+Clv:JLQCD,Nf2:DBW2+DW:RBC}.
We observe a reasonable agreement among the data suggesting that 
discretization errors are not large in this plot.
More importantly, 
\FIGURE{
   \centering
   \includegraphics[angle=0,width=0.45\linewidth,clip]{fPS_vs_MPS2.ar0.eps}
   \vspace{-3mm}
   \caption{
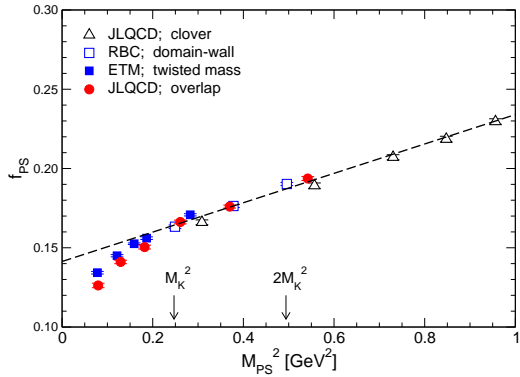
      Pion decay constant as a function of pion mass squared.
      Dashed line shows a linear fit to open symbols.
   }
   \label{fig:fK_fpi:fPS_vs_Mpi2}
}
\noindent 
data at $m_{ud, \rm sim} \lesssim m_s/2$ 
from recent simulations 
show a curvature toward the chiral limit as suggested 
by the chiral logarithm at NLO, 
whereas the curvature is not clear at heavier $m_{ud, \rm sim}$.
While data at small $m_{ud, \rm sim}$ are subject to effects due to finite lattice volume,
the ETM collaboration \cite{Nf2:Sym+tmW:ETM}
observe that their data with finite volume corrections \cite{FSE}
are described by the NLO ChPT formula reasonably well.
They obtain
\bea
   F = 121.3(7)~\mbox{MeV},
   \hspace{5mm}
   \bar{l}_4 = 4.52(6), 
\eea
which are consistent with lattice estimates of $F$ 
from the MILC's simulation in $p$-regime \cite{fK_fpi:Nf3:ImpG+ImpKS:MILC:2006:2} 
and JLQCD's one in $\epsilon$-regime \cite{e-regime:Nf2:RG+Ovr:JLQCD},
and with a phenomenological estimate of $\bar{l}_4$ \cite{l4}.
This suggests that 
recent simulations with Wilson-type and chiral fermions are now 
exploring $m_{ud, \rm sim}$
sufficiently small to make contact with NLO ChPT.

\FIGURE{
   \centering
   \includegraphics[angle=0,width=0.43\linewidth,clip]{fPS_compare.eps}
   \vspace{-3mm}
   \caption{
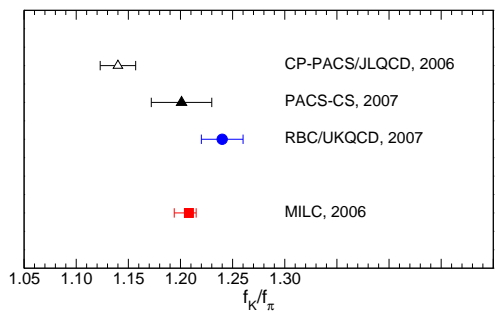
      Recent unquenched estimate of $f_K/f_\pi$.
   }
   \label{fig:fK_fpi:fPS}
}


Recent estimates of $f_K/f_\pi$ in three-flavor QCD 
are collected in Fig.~\ref{fig:fK_fpi:fPS}.
The CP-PACS and JLQCD collaborations obtain a slightly smaller result 
than others \cite{fK_fpi:Nf3:RG+Clv:CP-PACS+JLQCD},
probably because 
their simulations are limited 
to $m_{ud, \rm sim} \! \gtrsim \! m_s/2$ and 
$f_\pi$ is overestimated due to the lack of the chiral logarithm.
The PACS-CS collaboration employs the clover fermions with 
the L\"uscher's domain-decomposed HMC \cite{algorithm2}.
A good agreement of their \cite{fK_fpi:Nf3:RG+Clv:PACS-CS}  
and RBC/UKQCD's estimates \cite{Nf3:RG+DW:RBC+UKQCD}
%
%
with Eq.~(\ref{eqn:fK_fpi:MILC}) is very encouraging,
though their simulations are still on-going and/or the quoted 
error is statistical only.
These are not enough mature to be used to derive an world average,
and we simply quote Eq.~(\ref{eqn:fK_fpi:MILC}) 
as the current best estimate of $f_K/f_\pi$.
It is, however, worth emphasizing that 
estimates of $f_K/f_\pi$ are expected to be improved remarkably 
in the near future by on-going simulations with the Wilson-type and 
chiral fermions by various groups.

\section{$K_{l3}$ form factor} 
\label{sec:kl3}



The $K_{l3}$ decays provide a precise determination of $\Vus$ through
\bea
   \Gamma(K \to \pi l \bar{\nu}_l)
   & = & 
   \frac{G_\mu^2}{192\pi^3} M_K^5 \, C^2 \, I \, 
   \Vus^2 \, |f_+(0)|^2 \,  
   S_{\rm EW} \, (1 + \Delta_{\rm EM} + \Delta_{SU(2)}),
   \label{eqn:kl3:Vus}
\eea
where $C$ is the Clebsh-Gordon coefficient equal to 1 ($1/\sqrt{2}$) for 
neutral (charged) kaon decays.
The short- and long-distance radiative corrections, 
denoted by $S_{\rm EW}$ and $\Delta_{\rm EM}$, 
and $SU(2)$ breaking corrections $\Delta_{SU(2)}$ are theoretical inputs,
whereas the decay rate $\Gamma$ and the phase space integral $I$ are 
determined from experimental measurements.
The uncertainties in $\Vus$ due to these inputs are well below 1\%
\cite{Kl3:Flavianet}.

\begin{table}[t]
   \begin{center}
   \begin{tabular}{l|lllllllll}
   \hline
      \hline 
                       & $N_f$             & quark action      
                       & $a$[fm]           & $L$[fm] 
                       & $M_{\rm PS}$[MeV] \\
      \hline

      JLQCD \cite{Kl3:Nf2:Plq+Clv:JLQCD}
                       & 2                 & clover      
                       & 0.09              & 1.8
                       & $\gtrsim$ 600     \\
      RBC \cite{Kl3:Nf2:DBW2+DW:RBC}
                       & 2                 & domain-wall 
                       & 0.12              & 1.9
                       & $\gtrsim$ 490     \\
      \hline
      Fermilab/MILC/HPQCD \cite{Kl3:Nf3:impG+impKS:MILC+FNAL}
                       & 3                 & AsqTad {\footnotesize ($d$=clover)}
                       & 0.12              & 2.5
                       & $\gtrsim$ 500     \\
      RBC/UKQCD \cite{Kl3:Nf3:RG+DW:RBC+UKQCD}
                       & 3                 & domain-wall 
                       & 0.12              & 1.9, 2.9
                       & $\gtrsim$ 300     \\
      \hline 
   \end{tabular}
   \caption{
      Simulation parameters in unquenched lattice calculations of $f_+(0)$.
      The clover fermions are used for valence down quarks
      in Ref.\cite{Kl3:Nf3:impG+impKS:MILC+FNAL}.
   }
   \label{tbl:sim_param:unquenched}
   \end{center}
   \vspace{-7mm}
\end{table}


The dominant uncertainty of the present estimate of $\Vus$ 
therefore arises from theoretical determination of 
the normalization of the vector form factor $f_+(0)$
defined from the $K \! \to \! \pi$ matrix element
$   
   \langle \pi(p^\prime) | \bar{s} \gamma_\mu u | K(p) \rangle
   \! = \!
   (p+p^\prime)_\mu \, f_+(q^2) 
   +
   (p-p^\prime)_\mu \, f_-(q^2),
$
where $q^2 = (p-p^\prime)^2$.
The leading correction \cite{Kl3:f2} in the chiral expansion 
\bea
f_+(0) = 1 + f_2 + f_4 + O(p^6)
   \label{eqn:kl3:f0}  
\eea
is practically free of uncertainties ($f_2\!=\!-0.023$),
because any poorly known LECs do not appear in $f_2$
thanks to the Ademollo-Gatto theorem \cite{AG-theorem}.


However, the higher order correction $f_4$ contains LECs in 
the chiral Lagrangian both at $O(p^4)$ and $O(p^6)$.
A phenomenological estimate $f_4 \!=\! -0.016(8)$
based on the quark model was obtained by Leutwyler and Roos (LR) 
\cite{Kl3:f4:QM:LR},
and has been used in previous determinations of $\Vus$.
There has been remarkable progress in studies based on ChPT, 
where the evaluation of the tree-level contribution with 
LECs in the $O(p^6)$ Lagrangian is the most crucial issue
\cite{Kl3:f4:ChPT}.
Recent estimates ranging from $f_4\! = \!-0.007(9)$ to 
+0.007(12) are slightly larger than the LR estimate
due to a (partial) cancellation between loop and tree-level contributions.


\FIGURE{
   \centering
   \includegraphics[angle=0,width=0.45\linewidth,clip]{R14_RBC.eps}
   \vspace{-5mm}
   \caption{
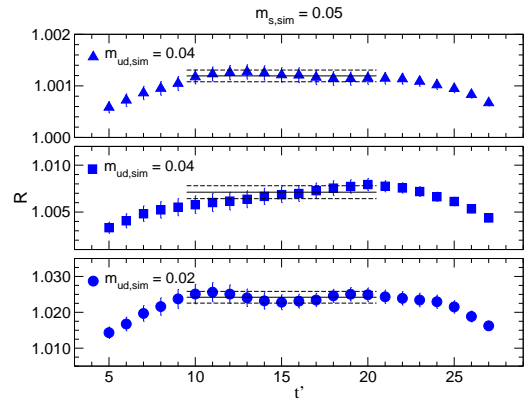
      Double ratio Eq.~(\ref{eqn:kl3:drat})
      as a function of $t^\prime$ with fixed $t\!=\!4$ and 
      $t^{\prime\prime}\!=\!28$.
      Data are from Ref.\cite{Kl3:Nf2:DBW2+DW:RBC}.
   }
   \label{fig:Kl3:drat}
   \vspace{3mm}
}

Lattice QCD can provide a {\it non-perturbative} determination of $f_+(0)$,
namely $f_4$ {\it and} higher order contributions.
Unquenched calculations performed so far are listed 
in Table~\ref{tbl:sim_param:unquenched}.
These studies basically follow the strategy proposed in the 
first calculation in quenched QCD \cite{Kl3:Nf0:Plq+Clv:BILMMSTV},
which is outlined below.


The first step is to calculate the scalar form factor 
$f_0 \! = \! f_+ + (q^2/(M_K^2-M_\pi^2))\,f_-$
from three point functions, {e.g.}
\bea
   C_\mu^{K \pi}(t,t^\prime,t^{\prime\prime})
   = 
   \langle 
      O_\pi(t^{\prime\prime}) | V_{\mu}(t^\prime) | O_K(t)^\dagger
   \rangle,
   \label{kl3:msn_corr_3pt}
\eea
where $O_{\pi(K)}(t)$ is the interpolation operator for pion 
(kaon) 
and $V_{\mu}(t)$ is the vector current
at the timeslice $t$.
With sufficiently large temporal separations 
$t^{\prime\prime}\!-\!t^{\prime}$ and $t^{\prime}\!-\!t$,
$C_\mu^{\pi K}(t,t^\prime,t^{\prime\prime})$ 
is dominated by the ground state contribution,
which is the matrix element $\langle \pi | V_{\mu} | K \rangle$ times 
unnecessary factors, such as 
the damping factor $e^{-M_K\,(t^{\prime}-t)}$.
These factors are canceled in the so-called double ratio 
\cite{double_ratio}. 
For instance, a double ratio
\bea
   \frac{ C_{4}^{K\pi}(t,t^\prime,t^{\prime\prime})\,
          C_{4}^{\pi K  }(t,t^\prime,t^{\prime\prime}) } 
        { C_{4}^{KK  }(t,t^\prime,t^{\prime\prime})\,
          C_{4}^{\pi \pi}(t,t^\prime,t^{\prime\prime}) }
   & \to & 
   \frac{(M_K+M_\pi)^2}{4 M_K M_\pi} |f_0(q_{\rm max})|^2
   \label{eqn:kl3:drat}
\eea
can be determined precisely as shown in Fig.~\ref{fig:Kl3:drat},
and it gives $f_0$ at $q_{\rm max}^2\!=\!(M_K\!-\!M_\pi)^2$ 
with an accuracy well below 1\%.
%
%
%
We can calculate $f_0$ at $q^2\!\ne\!q_{\rm max}^2$ from 
different double ratios proposed in Refs.~\cite{Kl3:Nf0:Plq+Clv:BILMMSTV,Kl3:Nf2:Plq+Clv:JLQCD,Kl3:Nf2:DBW2+DW:RBC},
which however 
involve three-point functions with nonzero meson momenta 
and hence are much noisier than the ratio Eq.~(\ref{eqn:kl3:drat}).



\FIGURE{
   \centering
   \includegraphics[angle=0,width=0.45\linewidth,clip]{lambda_vs_MPS2.eps}
   \vspace{-3mm}
   \caption{
      Lattice estimates of slope $\lambda_0$ 
      \cite{Kl3:Nf0:Plq+Clv:BILMMSTV,Kl3:Nf2:Plq+Clv:JLQCD,Kl3:Nf2:DBW2+DW:RBC,Kl3:Nf3:RG+DW:RBC+UKQCD}
      together with experimental values \cite{Kl3:lambda0:Exp}.
   }
   \label{fig:Kl3:lambda0}
}

Then, we interpolate $f_0$ to $q^2\!=\!0$.
%
The $q^2$ dependence is parametrized using 
the monopole ansatz $f_0(q^2)\!=\!f_0(0)/(1-\lambda_0\,q^2)$
or polynomial forms up to quadratic order
$f_0(q^2)\!=\!f_0(0) + \lambda_0\,q^2 
                     + \lambda_0^{\prime}\,q^4$.
These forms are also employed in analyses of experimental data.
It turns out that the choice of the interpolation form does not cause 
a large uncertainty in the unquenched studies,
since an accurate estimate of $f_0(q_{\rm max}^2)$ is available 
near the interpolation point $q^2\!=\!0$.
It is also encouraging to observe in Fig.~\ref{fig:Kl3:lambda0}
that $\lambda_0$ from lattice studies 
shows a reasonable agreement with experimental measurements
\cite{Kl3:Flavianet}.

\FIGURE{
   \centering
   \includegraphics[angle=0,width=0.45\linewidth,clip]{f0_vs_MPS2.nf3.eps}
   \vspace{-18mm}
   \caption{
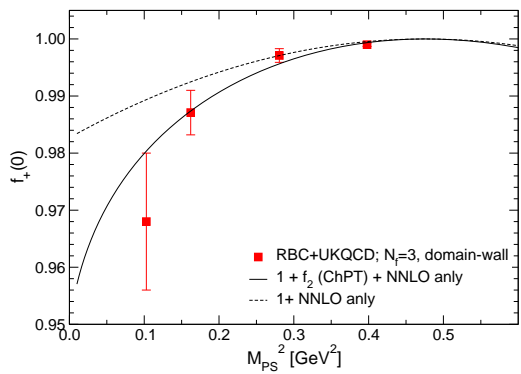
      Vector from factor as a function of $m_{ud, \rm sim}$.
      The solid line shows a fit curve Eq.~(\ref{eqn:kl3:chiral_fit:linear}).
      The NLO term $f_2$ is subtracted in the dashed line; namely
      the difference between two lines shows $f_2$.
   }
   \label{fig:kl3:mq_dep}
}


Finally, $f_+(0)\!=\!f_0(0)$ is extrapolated to 
the physical quark masses $m_{ud}$ and $m_s$.
In all unquenched calculations, 
a ratio motivated by the Ademollo-Gatto theorem
\bea
   R = \frac{f_+(0)-1-f_2}{(M_K^2-M_\pi^2)^2}
   \label{eqn:kl3:AGratio}
\eea
can be fitted to a rather simple polynomial form 
\bea
R = c_0 + c_1\,(M_K^2+M_\pi^2).
   \label{eqn:kl3:chiral_fit:linear}
\eea
It is possible that,
since most simulations are limited to heavy quark masses
$m_{ud, \rm sim} \! \gtrsim \! m_s/2$, 
the NNLO (and higher order) chiral logs vary smoothly in this region 
and are well approximated by the analytic terms.

In order to get an idea about how small $m_{ud, \rm sim}$ is needed
to see the chiral logs in $f_+(0)$ clearly,
data from the RBC/UKQCD's study is plotted 
as a function of $m_{ud,\rm sim}$
in Fig.~\ref{fig:kl3:mq_dep}.
The NLO chiral log $f_2$ rapidly increases at $m_{ud,\rm sim} \lesssim m_s/2$.
This suggests that precise lattice data in this region are essential
for a reliable chiral extrapolation 
compatible with the existence of the chiral logs.

We note that 
the error of $f_+(0)$ may rapidly increase 
with decreasing $m_{ud, \rm sim}$,
because of longer auto-correlations of gauge configurations 
and larger $q_{\rm max}^2$ for the $q^2$ interpolation of $f_0(q^2)$.
%
%
%
%
In future studies at small $m_{ud,\rm sim}$, therefore, 
it is advisable to employ improved measurement methods,
such as the all-to-all quark propagators
to improve the accuracy of $f_0(q^2)$ \cite{Kl3:A2A}.
The twisted boundary condition,
which enables us to explore $q^2 \! \sim \! 0$ \cite{Kl3:TBC},
and a model independent parametrization of the $q^2$ dependence of $f_0$
\cite{Disp_bound} are useful to reduce systematic uncertainties due to
the $q^2$ interpolation.


Figure~\ref{fig:Kl3:f0} shows recent lattice estimates of $f_+(0)$.
The RBC/UKQCD collaboration confirms that 
finite volume corrections at $L \! \sim \! 2$~fm are small 
down to $m_{ud, \rm sim} \! \approx \! m_s/4$.
This observation is encouraging since other unquenched studies 
are conducted with similar or larger lattice sizes.
The nice consistency among lattice results 
may suggest that discretization and quenching errors are not large.

\FIGURE{
   \centering
   \includegraphics[angle=0,width=0.45\linewidth,clip]{f0_compare.eps}
   \vspace{-5mm}
   \caption{
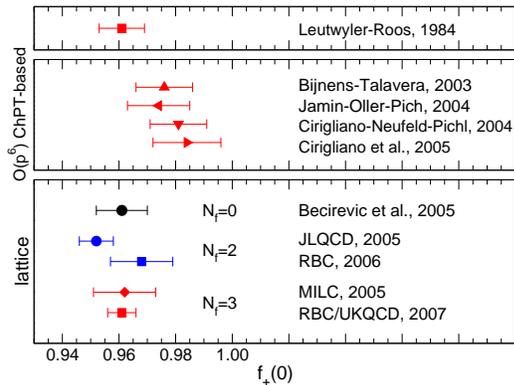
      Vector form factor $f_+(0)$ from phenomenological model 
      (top panel), ChPT (middle panel) and lattice QCD (bottom panel).
   }
   \label{fig:Kl3:f0}
}

All lattice results are in good agreement with the LR value. 
We note that, however, estimates from ChPT are slightly 
higher due to the NNLO loop contributions.
Therefore,
the agreement between lattice and the LR value has to be examined
carefully by precise lattice calculations at 
$m_{ud, \rm sim} \! \lesssim \! m_s/2$, 
where chiral logs are expected to be seen clearly as discussed above.

\section{Conclusions} 
\label{sec:summary}
\vspace{-1mm}

From the MILC's estimate of $f_K/f_\pi$ and the $K_{\mu 2}$/$\pi_{\mu 2}$
decay rates, we obtain $\Vus\!=\!0.2226$ $(+26/\!-\!15)$.
The preliminary result of $f_+(0)\!=\!0.9609(51)$ 
from the RBC/UKQCD's study
and $|V_{\rm us}\,f_+(0)|\!=\!0.21673(46)$ 
from the FlaviaNet working group \cite{Kl3:Flavianet} lead to
$\Vus\!=\!0.2255(13)$ 
which is consistent with the value quoted earlier.
The latter estimate needs, however, 
further studies to increase 
the reliability of the chiral extrapolation of $f_+(0)$.

For both $f_K/f_\pi$ and $f_+(0)$,
we observe that precise data at sufficiently small $ud$ quark masses,
typically $m_{ud, \rm sim} \! \lesssim \! m_s/2$,
are needed for a reliable chiral extrapolation.
%
%
Thanks to recent algorithmic improvements,
several groups have already started large-scale simulations 
in this region of $m_{ud, \rm sim}$
with different fermion discretizations.
While their results are premature to be taken into account 
in the above estimates of $\Vus$,
lattice estimates of $f_K/f_\pi$ and $K_{l3}$ form factors 
are expected to be remarkably improved by these studies 
in the {\it near} future \cite{Lat07:Juettner}.

\section{Acknowledgments}
\vspace{-1mm}

I thank S.~Hashimoto, A.~J\"uttner and J.~Noaki for helpful correspondence. 
This work is supported by the Grant-in-Aid of the Ministry of Education, 
Culture, Sports, Science and Technology of Japan (No.~17740171).

\end{document}